\documentclass[12pt]{report}
\usepackage{amsmath,latexsym}

\renewcommand{\theequation}{\arabic{equation}}

\def\ds{\displaystyle}

\begin{document}

\title{Differential Geometrical Methods \\
for Deriving Dirac's Equation \\ 
in Curved Spacetime \\ to Account for the Presence of Matter}
\author{Joseph Saaty, PhD \\
The Union Institutute \\
School of Interdisciplinary Arts and Sciences} 

\maketitle

\baselinestretch{2}
\baselineskip=22pt

\def\theequation{\arabic{equation}}
\setcounter{equation}{0}

\begin{abstract}
Differential geomtrical methods for deriving the Dirac equation 
in Curved Spacetime are presented.
Einstein's field equation is applied in a novel manner; in the most 
current standard reference, Birrell and Davies, 1994 [1], 
the suggestions for deriving the Dirac equation in Curved Spacetime 
make no mention of employing Einstein's field equation. 
Thus, to date, the literature on the derivation of the Dirac equation
could not include an expression for the presence of matter.  This lack is 
consistent with earlier publications, including Lichnerowicz's well-known 
1964 journal article [3], which presented the first such derivation, and 
Dimock's 1982 article [2].  The new differential geometrical methods 
go beyond all previous suggestions, which only apply to cases in the 
absence of matter.  These differential geometrical methods have 
resulted in derivations of the Dirac equation in Curved Spacetime 
that apply to either the presence or absence of matter.

\end{abstract}


General Relativity had been initiated by Einstein through establishing 
an interdisciplinary contact between differential geometry
and Special Relativity, where differential geometry contains the  
concept of curved space and Special Relativity contains the concept
of time. This cross-fertilization dramatically introduced
the concept of Curved Spacetime.
Einstein's cross-fertilization brought forward the concept of Curved Spacetime 
in the juncture of differential geometry and relativistic field theory.
The research I present here follows this same combinatory vein of thought 
by bringing together differential geometry and relativistic \emph{quantum} 
field theory, adding to Curved Spacetime the
concept of spin quantum number.  This combination of the
most advanced tools available, from both General Relativity and
quantum field theory, will allow us to perform investigations
at the frontiers of science.

In this article, the techniques I employ are a part of my original
research; however, I was inspired by the work of Lichnerowicz,  
and I further developed
mathematical leads suggested in his paper \cite{3}.

Let us begin by making the transition from Flat to Curved Spacetime.\\

If $\eta_{\alpha\beta}$ denotes the Minkowskian metric tensor (flat 
spacetime) and if $g_{\mu\nu}$ denotes the Riemannian metric tensor (curved 
spacetime), then $g_{\mu\nu}(x) = V^{\alpha}_{\mu}(x) V^{\beta}_{\nu}(x)
\eta_{\alpha\beta}$,  \ $\eta_{\alpha\beta} =$ diag(1, -1, -1, -1).
If $\gamma^{\alpha}$ denotes Dirac's Spinor matrix in flat spacetime \cite{3}
$$
\gamma^{\mu} = V^{\mu}_{\alpha} \gamma^{\alpha} \ \ \mbox{will be the} \
\ \gamma^{\alpha}
\ \  \mbox{version in curved spacetime.} \ \ V_{\alpha} 
= V^{\beta}\eta_{\alpha\beta}. \leqno{\textrm{therefore}}
$$
$$
\frac 12 V^{\alpha} V^{\beta}(\gamma_{\alpha}\gamma_{\beta} + 
\gamma_{\beta}\gamma_{\alpha})  = V^{\alpha}V^{\beta}\eta_{\alpha\beta}
$$

\medskip

\noindent then 
$\ds \gamma_{\alpha}\gamma_{\beta} + \gamma_{\beta}\gamma_{\alpha} = 
2\eta_{\alpha\beta}$  which shows the anti-commutation relation still hold 
or 
$
      \gamma^{\lambda}\gamma^{\mu} + \gamma^{\mu}\gamma^{\lambda} =
2\eta^{\mu\lambda} = 2\eta^{\lambda\mu}.
$
Let 
$\ds \Delta = \gamma^{\lambda}\gamma^{\mu}
\nabla_{\lambda} \nabla_{\mu}$
where $\nabla$ is the $\partial$ operaor version in curved spacetime so that 
$\partial$ in flat spacetime will transform to $\nabla$ where
$\nabla_b \omega_c = \partial_b \omega_c - \Gamma^d_{bc} \omega_d$, \ 
$\Gamma^d_{bc}$ is the Christoffel symbol.
\begin{align*}
   \Delta\Psi &= \frac 12 \underbrace{(\gamma^{\lambda}\gamma^{\mu} + 
\gamma^{\mu}\gamma^{\lambda})}_{2\eta^{\lambda\mu}} \nabla_{\lambda} 
\nabla_{\mu} \Psi + \frac 12 (\gamma^{\lambda}\gamma^{\mu} - 
\gamma^{\mu}\gamma^{\lambda}) \nabla_{\lambda} \nabla_{\mu} \Psi\\
\intertext{so that the above equation can be written as}
   \Delta\Psi  &= \eta^{\lambda\mu} \nabla_{\lambda} \nabla_{\mu} \Psi + 
\frac 12(\gamma^{\lambda}\gamma^{\mu} - 
\gamma^{\mu}\gamma^{\lambda}) \nabla_{\lambda} \nabla_{\mu} \Psi \\ 
   \Delta\Psi  &= \nabla^{\lambda} \nabla_{\lambda} \Psi + 
\frac 12 \gamma^{\lambda}\gamma^{\mu} 
\underbrace{(\nabla_{\lambda}\nabla_{\mu} - 
\nabla_{\mu}\nabla_{\lambda})}_{[\nabla_{\lambda},\nabla_{\mu}]} \Psi
\end{align*}
Thanks are due to Lichnerowicz for having brought us thus far.
I will now introduce my own extension of this method by 
involving Einstein's field equation in order to account for the
presence of matter.

The following three equations from General Relativity will be used \cite{4}
\begin{itemize}
\item[(i)] $[\nabla_{\lambda},\nabla_{\mu}] V^{\ell} = R^{\ell}_{\nu\lambda\mu}V^{\nu}$, \ \ 
where $R^{\ell}_{\nu\lambda\mu}$ is the Riemann Curvature tensor.
\item[(ii)] $R_{\mu\nu} - \dfrac 12 g_{\mu\nu} R = 0$ \ Einstein's field 
equation in absence of any matter.
\item[(iii)] $R_{\mu\nu} - \dfrac 12 g_{\mu\nu} R + kT_{\mu\nu} = 0$ \ Einstein's 
field equation in presence of matter.
\end{itemize}
$$
  \Delta\Psi = \nabla^{\lambda}\nabla_{\lambda} \Psi + \frac 12 
\gamma^{\lambda}\gamma^{\mu} [\nabla_{\lambda},\nabla_{\mu}] \Psi.
$$
We notice that from (i) the last term to the right side of the above equation 
can be expressed in terms of the Riemann Curvature tensor and thus we can 
conclude that Dirac's equation in curved spacetime gave rise to an extra term 
depends on $R$.

To find the value of $\dfrac 12 \gamma^{\lambda}\gamma^{\mu}
[\nabla_{\lambda},\nabla_{\mu}] \Psi$ in terms of $R$.
\begin{align*}
\frac 12 \gamma^{\lambda}\gamma^{\mu} [\nabla_{\lambda},\nabla_{\mu}] \Psi^{\ell} 
&= \frac 12 \gamma^{\lambda}\gamma^{\mu} R_{\nu\lambda\mu}^{\ell}\Psi^{\nu} \\
&= \frac 12 \gamma^{\lambda}\gamma^{\mu} R_{\nu\lambda\mu}^{\nu}\Psi^{\ell} \\
&= \frac 12 \gamma^{\lambda}\gamma^{\mu} R_{\lambda\mu} \Psi^{\ell}.
\end{align*}
Since $R_{\nu\lambda\mu}^{\nu} = R_{\lambda\mu}$, the above equation
can be written as
\begin{align*}
\frac 12 \gamma^{\lambda}\gamma^{\mu} [\nabla_{\lambda},\nabla_{\mu}] \Psi^{\ell} 
&= \frac 12 \gamma^{\lambda}\gamma^{\mu} R_{\lambda\mu}\Psi^{\ell} \\
\frac 12 \gamma^{\lambda}\gamma^{\mu} [\nabla_{\lambda},\nabla_{\mu}] \Psi 
&= \frac 12 \gamma^{\lambda}\gamma^{\mu} \underbrace{R_{\lambda\mu}}_{\frac 12  
g_{\lambda\mu}R} \Psi.
\end{align*}
But (ii) is $R_{\lambda\mu} =  \dfrac 12 g_{\lambda\mu}R$ 
Einstein's field equation in absence of any matter. 

\medskip

\noindent Therefore 
$\dfrac 12 \gamma^{\lambda}\gamma^{\mu} [\nabla_{\lambda},\nabla_{\mu}] 
\Psi = \dfrac{1}{4} \gamma^{\lambda}
\gamma^{\mu}g_{\lambda\mu}R \Psi$.  
But $\gamma^{\mu}g_{\lambda\mu} = \fbox{$\gamma^{\mu}g_{\mu\lambda} = 
\gamma_{\lambda}$}\ , \ \ g_{\mu\lambda} = g_{\lambda\mu}$
$$
  \frac 12 \gamma^{\lambda}\gamma^{\mu} [\nabla_{\lambda},\nabla_{\mu}] 
 \Psi = \frac 14 \gamma^{\lambda}\gamma_{\lambda} R\Psi \leqno{\textrm{so that}}
$$
$$
  \sum^{4}_{\lambda = 1} \gamma^{\lambda}\gamma_{\lambda} = 4 \quad 
    \mbox{each of} \quad \gamma^{\lambda}\gamma_{\lambda} = 1
$$
$$
     \sum_{\mu} \dfrac 12 \gamma^{\lambda}\gamma^{\mu}
       [\nabla_{\lambda},\nabla_{\mu}] \Psi = \dfrac 14 
\underbrace{(\gamma^{\lambda}\gamma_{\lambda})}_{1} R\Psi
\leqno{\textrm{and}}
$$
becomes,
\begin{align*}
\sum^{}_{\mu} \frac 12 
\gamma^{\lambda}\gamma^{\mu} [\nabla_{\lambda},\nabla_{\mu}] \Psi 
&= \frac 14 (\gamma^{\lambda}\gamma_{\lambda})R\Psi, 
\quad \lambda,\mu = 1,2,3,4 \ \ \ \mbox{implies that} \\
\sum^{}_{\mu} \frac 12 
\gamma^{1}\gamma^{\mu} [\nabla_{1},\nabla_{\mu}] \Psi 
&= \frac 14 \underbrace{(\gamma^{1}\gamma_{1})}_{1} R\Psi, \\
\sum^{}_{\mu} \frac 12 
\gamma^{2}\gamma^{\mu} [\nabla_{2},\nabla_{\mu}] \Psi 
&= \frac 14 \underbrace{(\gamma^{2}\gamma_{2})}_{1} R\Psi, \\
\sum^{}_{\mu} \frac 12 
\gamma^{3}\gamma^{\mu} [\nabla_{3},\nabla_{\mu}] \Psi 
&= \frac 14 \underbrace{(\gamma^{3}\gamma_{3})}_{1} R\Psi, \\
\sum^{}_{\mu} \frac 12 
\gamma^{4}\gamma^{\mu} [\nabla_{4},\nabla_{\mu}] \Psi 
&= \frac 14 \underbrace{(\gamma^{4}\gamma_{4})}_{1} R\Psi
\end{align*}

The right side of the above equations are the same, i.e., $\dfrac 14 R\Psi$, 
therefore the left side of the above four equations are equal to 
each other. Thus, the addition of the four left sides $= 4 \times$ (anyone 
of them) $= 4 \ds\sum^{}_{\mu} \dfrac 12 \gamma^{\lambda}\gamma^{\mu} 
[\nabla_{\lambda},\nabla_{\mu}] \Psi$,  \ $\lambda$ can be any of 1, 2, 3, or 
4, but the addition of the four left sides $=$ the addition of the four right 
sides
\begin{align*}
4 &\sum^{}_{\mu} \frac 12 
\gamma^{\lambda}\gamma^{\mu} [\nabla_{\lambda},\nabla_{\mu}] \Psi = R\Psi, \\
&\sum^{}_{\mu} \frac 12 
\gamma^{\lambda}\gamma^{\mu} [\nabla_{\lambda},\nabla_{\mu}] \Psi = 
\frac 14 R\Psi, 
\end{align*}
or $\ds \sum^{}_{\mu} \dfrac 12 \gamma^{\lambda}\gamma^{\mu} 
[\nabla_{\lambda},\nabla_{\mu}] \Psi = \dfrac 14 R\Psi$.
Therefore, $\Delta\Psi = \nabla^{\lambda}\nabla_{\lambda}\Psi + \dfrac 14 
R\Psi$.

In absence of any matter, Dirac's equation in flat spacetime 
$(\gamma^{\mu}\gamma^{\nu}\partial_{\mu}\partial_{\nu} + m^2) \Psi = 0$ will 
become
\begin{equation*}
\fbox{$(\nabla^{\lambda}\nabla_{\lambda} + \dfrac 14 R + m^2) \Psi = 0$}
\ \ \mbox{in curved spacetime}
\end{equation*}
However, in presence of matter,  from (iii) we get for 
Dirac's equation in curved spacetime
$$
\fbox{$(\nabla^{\lambda}\nabla_{\lambda} + \dfrac 14 R -
\dfrac 12 K \gamma^{\lambda}\gamma^{\mu} T_{\lambda\mu} + m^2) \Psi = 0$}
$$

\noindent Thus we were able to get new information of physical significance, and
that is, Einstein's field equation in presence of matter will lead to 
Dirac's equation in curved spacetime in presence of matter which is
$$
  (\nabla^{\lambda}\nabla_{\lambda} + \frac 14 R -
     \frac 12 K\gamma^{\lambda}\gamma^{\mu} T_{\lambda\mu} + m^2)\Psi  = 0.
$$

My method of deriving the above two equations provided new information
as far as the Dirac equation in curved spacetime is concerned.  This
is due to the fact that Einstein's equations are separated.  One due
to the absence of matter and the other, due to the presence
of matter; however, this has not been presented in books and journals
relating to quantum field theory in curved spacetime.

ACKNOWLEDGEMENT:

The author would like to express his gratitude to theoretical physicist
Professor Alan Chodos of Yale University for his encouragement, particularly
for motivating me to pursue my line of thinking in curved spacetime.

\bibliographystyle{amsplain}

\end{document}